\documentclass[amsmath,amssymb,aip,jcp,twocolumn,10pt]{revtex4-1}

\usepackage{dcolumn}
\usepackage{graphicx}
\usepackage{graphics}
\usepackage{setspace}
\usepackage[usenames]{color}

\newcommand{\tH}{\tilde{H}}
\newcommand{\hH}{\hat{H}}
\newcommand{\hV}{\hat{V}}
\newcommand{\bv}{\bar{v}}
\newcommand{\bH}{\bar{H}}
\newcommand{\cH}{{\cal H}}

\newcommand{\hT}{\hat{T}}
\newcommand{\hR}{\hat{R}}
\newcommand{\hLe}{\hat{L}}
\newcommand{\hF}{\hat{F}}
\newcommand{\hW}{\hat{W}}
\newcommand{\bra}[1]{\langle #1|}
\newcommand{\ket}[1]{| #1\rangle}
\newcommand{\rmx}{{\rm x}}
\newcommand{\rmy}{{\rm y}}

\begin{document}

\title{Computing solvated excited states using fragment-effective-field coupled-cluster perturbation theory with application to the electronic spectra of nucleobases in water}

\author{Jason N. Byrd}
\email{byrd.jason@ensco.com}
\affiliation{ENSCO, Inc., 4849 North Wickham Road, Melbourne, Florida 32940, USA}
\affiliation{Quantum Theory Project, University of Florida, Gainesville, FL 32611}
\author{Robert W. {Molt, Jr.}}
\author{Rodney J. Bartlett}
\affiliation{Quantum Theory Project, University of Florida, Gainesville, FL 32611}
\author{Beverly A. Sanders}
\affiliation{Quantum Theory Project, University of Florida, Gainesville, FL 32611}
\affiliation{Department of Computer and Information Science and Engineering, University of
Florida, Gainesville, FL 32611}

% \begin{spacing}{2.0}

\begin{abstract}
DRAFT
\end{abstract}

\maketitle

\section{Introduction}

Spectroscopy of molecules in a crystal or liquid solution is a standard experimental scenario with little theoretical support.  Most computational techniques currently available are impractical for use on molecular systems beyond a few dozen atoms.  This restriction to only dozens of atoms is a serious difficulty because bulk limit properties only stabilize in the nanometer length scale with a thousand or more molecules.

% {\color{blue}Motivate using solvated nucleobases.}

A powerful technique used in studying large molecular systems\cite{akimov2015,collins2015,raghavachari2015} is to fragment the system into individual units that are small enough to be treated using standard techniques with the surrounding system treated as an interacting bath.  Molecular systems that have a low degree of delocalization are optimal candidates for a fragmentation approach to solving the Schrödinger equation.  Liquids or non-covalently bonded molecular crystals are two general examples that typically fit in the distinct unit category.  Most quantum mechanical fragmentation techniques begin with the same Born-Oppenheimer antisymmetrized product wavefunction Ansatz, 
\begin{equation}%\label{Aprodspace}
\ket{\Psi} = \hat{A}\ket{\phi^1}\otimes\ket{\phi^2}\otimes\dots\ket{\phi^n},
\end{equation}
and proceed to build upon this mean-field with various kinds of approximations {\color{blue}n-body expansion, embedding, dimers in molecules, etc.}
Our fragment approach to solving the fragmented Schr\"{o}dinger equation, based on coupled-cluster perturbation theory, is a linear-scaling method (w.r.t number of fragments) that generates a highly structured and sparse representation of the entire physical system while retaining many of the desirable aspects of full perturbation theory.

{\color{blue}Fragment-Effective-Field as a way of doing CCPT with relaxed pairs.  New in this approach is an alternative mean-field to MCPT where the result is canonical and includes full relaxation.}

The aim of this work is to show that nanoscale spectroscopy in liquids is obtainable with our FEF-CCPT method, where excited states belonging to the fragment of interest can be computed directly from the FEF effective Hamiltonian using perturbation theory.  Excited states computed in this way will include all electrostatic and polarization effects from the surrounding bulk\footnote{Bulk is defined to be when the interior properties of a cluster are unchanged by the addition of a molecule to the surface of the cluster.} material, making it a good approximation for systems with negligible excited state energy transfer between fragments.  Using the FEF-CCPt excited state method it is then possible to provide guidance to experimental efforts on the long-standing problems in biochemistry.

\section{\label{theory}Theory}

\subsection{Fragment Mean Field}

In a fragment based theory the total antisymmetric Hilbert space, $\cH$, is partitioned into $n$ fragment specific antisymmetric subspaces, $\cH_\rmx$, with 
\begin{equation}
\cH = \cH_1 \otimes \cH_2 \otimes \dots \otimes \cH_n.
\end{equation}
The antisymmetric wavefunction spanning this product space is 
\begin{equation}\label{Aprodspace}
\ket{\Psi} = \hat{A}\ket{\phi^1}\otimes\ket{\phi^2}\otimes\dots\ket{\phi^n}
\end{equation}
where each subspace $\ket{\phi^\rmx}$ is itself antisymmetric and $\hat{A}$ is the total antisymmetrization operator.  
To facilitate a second-quantized representation we define the following monomer specific index domains
\begin{equation}
p,p_1,r,r_1,a,i,a_1,i_1,\dots\in \lbrace\cH_\rmx\rbrace
\end{equation}
and
\begin{equation}
q,q_1,s,s_1,b,j,b_1,j_1,\dots\in \lbrace\cH_\rmy\rbrace
\end{equation}
where the indices $i,j,\dots$ and $a,b,\dots$ refer to occupied and virtual
molecular orbitals (MO) respectively, while $p=i\cup a,~q=j\cup b,~\dots$ ranges
over both occupied and virtual.  Assuming a monomer centered basis set, the
atomic orbital indices can also be broken into subspaces as
\begin{equation}
\mu,\mu_1,\nu,\nu_1,\dots\in\lbrace\cH_\rmx\rbrace
\end{equation}
and
\begin{equation}
\lambda,\lambda_1,\sigma,\sigma_1,\dots\in\lbrace\cH_\rmy\rbrace.
\end{equation}  

The mean field wavefunction Ansatz central to many fragment approaches is the direct product space of antisymmetric fragment wavefunctions,
\begin{equation}\label{prodspace}
\ket{\Psi} = \ket{\phi^1}\otimes\ket{\phi^2}\otimes\dots\ket{\phi^n},
\end{equation}
where the approximation $\hat{A}\sim 1$ has been made.  The fragment wavefunctions are chosen to satisfy the Hartree-Fock (SCF) equation and serve as that fragment's mean-field solution, with the associated monomer centered Gaussian basis functions $\chi_\mu({\bf r}_1)$ and Hartree-Fock coefficients $C_{\mu p}$.  As a consequence of the monomer centered basis set, the one- and two-particle electronic Hamiltonian can only have elements belonging to at most two fragments.  Within the $\hat{A}\sim 1$ approximation there are two clear choices in mean-field choice.  First is the non-interacting SCF solution obtained with the fragment Fock matrix
\begin{equation}
{\cal\bf F} = {\bf h}(1) + (2{\bf J} - {\bf K})
\end{equation}
with inter-fragment contributions included as a perturbation.  The second mean-field choice is to include the inter-fragment contributions directly into the Fock matrix so as to introduce inter-fragment polarization screening
\begin{equation}
\tilde{\cal\bf F} = {\bf h}(1) + (2{\bf J} - {\bf K}) + {\pmb \omega}_{\rmx}.
\end{equation}
Here the one-particle operator\cite{rybak1991} 
\begin{equation}\label{wc}
{\pmb \omega}_{\rmx}=\tilde{h}^{\mu}_{\nu} + 2*v^{\mu j}_{\nu j}
\end{equation}
is composed of the inter-fragment one-electron integrals
\begin{equation}
\tilde{h}^{\mu}_{\nu} = 
-\int d{\bf r_1}d{\bf r_2}
\chi^*_{\mu}({\bf r_1}) 
\sum_{\substack{\alpha\in \rmy\\ \rmy\ne\rmx}} \frac{Z_\alpha}{R_\alpha-r_1}
\chi_{\nu}({\bf r_1})
\end{equation}
which includes the electrostatic contribution of all nuclei in the system on $\rmx$.  
The two-electron "bubble diagram" integrals are evaluated from the standard
expression
\begin{multline}\label{dp2e}
v^{\mu j}_{\nu j} = 
\sum_j\sum_{\lambda\sigma}
    C_{\lambda j}C_{\sigma j}
    \times
    \\
\int d{\bf r_1}d{\bf r_2}
\psi^*_{\mu}({\bf r_1}) \psi_{\nu}({\bf r_1}) 
r^{-1}_{12}
\psi^*_{\lambda}({\bf r_2}) \psi_{\sigma}({\bf r_2}).
\end{multline}
{\color{blue}same as solving CPHF iterativly = analogous to RPA GW.  Compare with FMO-2 mean-field.}

%{\color{red}Evidence suggests the causes of spurious states arises from over polarization due to %fragment-centered basis set fragmentation.\cite{horn2015}  Further exchange based attenuation %possible.
%
%Alternative is to use something akin to divide and conquer SCF or other related linear scaling %local SCF methods.
%}

The non-interacting was our previous choice for the MCPT method\cite{byrd2015-a} and has the benefit of a simple and well defined wavefunction, with all inter-fragment contributions included in the post-SCF perturbation theory.  A disadvantage to the non-interacting field is that the ${\pmb \omega}_{\rmx}$ is constructed using the non-perturbed SCF density from fragment $\rmy$, and so any relaxation in the one-electron part of the wavefunction has to come from high-orders of perturbation theory.  The alternative polarization-screened SCF solution we propose here has the advantage that the ${\pmb \omega}_{\rmx}$ is fully relaxed to infinite-order, with the added benefit that the final SCF solution is canonical which provides simplifications in the final expressions.  It should be noted that in either mean-field choice, the Hartree-Fock orbitals on each fragment are not orthogonalized with other fragments necessitating careful use of the ${\bf S}$ overlap matrix. 

Using the polarization-screened mean-field molecular orbital basis, the normal ordered, second-quantized Hamiltonian is then partitioned into one- and two-monomer contributions, where
\begin{multline}\label{NHx}
\hH_\rmx =
\langle\hH_\rmx\rangle^\rmx +
\sum_{pp_1}f^p_{p_1} \lbrace \hat{p}^\dag \hat{p}_1\rbrace +  
\frac{1}{4}\sum_{pp_1rr_1}\bar{v}^{pp_1}_{rr_1} \lbrace 
\hat{p}^\dag \hat{p}^\dag_1 \hat{r}_1\hat{r}\rbrace 
\end{multline}
and
\begin{equation}\label{NHxy}
\hat{H}_{\rmx\rmy} =  
\langle\hH_{\rmx\rmy}\rangle^{\rmx\rmy}
+ \sum_{pqrs}{v}^{pq}_{rs} \lbrace \hat{p}^\dag \hat{q}^\dag \hat{s}\hat{r}\rbrace,
\end{equation}
where $f$ is the usual one-particle canonical Fock matrix, $v$ ($\bar{v}$) are (anti-)symmetric two-electron integrals and $\lbrace\rbrace$ denotes normal ordering of the included operators. As a simplification we use the shorthand
\begin{equation}
\langle\cdots\rangle^{\rmx\rmy} = 
\bra{\phi^{\rmx\rmy}_0}\cdots\ket{\phi^{\rmx\rmy}_0}=
\bra{\phi^{\rmx}_0}\bra{\phi^{\rmy}_0} \cdots \ket{\phi^{\rmx}_0}\ket{\phi^{\rmy}_0}
\end{equation}
and denote the two-particle operators as
\begin{equation}
\hW_\rmx = \frac{1}{4}\sum_{pp_1rr_1}\bar{v}^{pp_1}_{rr_1} \lbrace \hat{p}^\dag \hat{p}^\dag_1 \hat{r}_1\hat{r}\rbrace
\end{equation}
and
\begin{equation}
\hW_{\rmx\rmy} = \sum_{pqrs}{v}^{pq}_{rs} \lbrace \hat{p}^\dag \hat{q}^\dag \hat{s}\hat{r}\rbrace.
\end{equation}
The system Hamiltonian can now be compactly expressed as
\begin{equation}  
\label{clusterH}
\hH = \sum^N_{\rm x} \hH_{\rm x} + \sum^N_{\rm x,y\ne x} \hH_{\rm xy}.
\end{equation}
with
\begin{equation}
\hH_\rmx = \langle\hH_\rmx\rangle^\rmx + \hF_\rmx + \hW_\rmx
\end{equation}
and 
\begin{equation}
\hH_{\rmx\rmy} = \langle\hH_{\rmx\rmy}\rangle^{\rmx\rmy} + \hW_{\rmx\rmy}.
\end{equation}

\subsection{Effective Hamiltonian Representation}

Employing an exponential cluster operator Ansatz, $e^{\hT}\ket\Psi$, on the product wavefunction Eq.  \ref{prodspace} requires partitioning the cluster operator into fragment subspaces.  Given a $g$-particle cluster operator $\hT_g$, it is clear that $\hT$ includes possible excitations from $1,\dots,g$ fragments (but never more than $g$).  This means that $\hT$ can be expressed as
\begin{equation}  \label{clusterT}
\hT = \sum_\rmx \hT^\rmx + \sum_{\rmx,\rmy\ne \rmx} \hT^{\rmx\rmy} + \dots
\end{equation}
with 
\begin{equation}
\hT^\rmx = \hT^\rmx_1+\hT^\rmx_2+\dots ~\text{and}~ \hT^{\rmx\rmy} =\hT^{\rmx\rmy}_1 +\hT^{\rmx\rmy}_2 +\dots.
\end{equation}  
The single excitation cluster operators are
\begin{equation} \label{1mt1}
\hT^\rmx_1 = \sum_{ai} t^a_i \hat{a}^{\dagger} \hat{i}
\end{equation}
and
\begin{equation}
\hT^{\rmx\rmy}_1 = 0
\end{equation}
(there can be no $\hT^{\rmx\rmy}_1$ operator in a monomer centered basis set), while the doubles cluster operators are
\begin{equation}  \label{1mt2}
\hT^\rmx_2  = \frac{1}{4}\sum_{aa_1ii_1} t^{a a_1}_{i i_1} \hat{a}^{\dagger} \hat{a}_1^{\dagger}\hat{i}_1 \hat{i},
\end{equation} 
and
\begin{equation}  \label{1dt2}
\hT^{\rmx\rmy}_2 = \sum_{abij}t^{a b}_{i j} \hat{a}^{\dagger} \hat{b}^{\dagger} \hat{j}
\hat{i}.
\end{equation}  

The coupled-cluster similarity transformed Hamiltonian,
\begin{equation}\label{hbar}
\bH \equiv e^{-\hT}\hH e^{\hT},
\end{equation}
is upon expanding Eqs. \ref{clusterH} and \ref{clusterT}
\begin{align}\label{fef2hbar1}
\bH &= 
e^{-\hT^\rmx}e^{-\hT^\rmy}e^{-\hT^{\rmx\rmy}}
(\hH_{\rmx}+\hH_{\rmy}+\hH_{\rmx\rmy})
e^{\hT^{\rmx\rmy}}e^{\hT^\rmy}e^{\hT^\rmx} \\
&=\label{fef2hbar2}
e^{-\hT^\rmx}\tH_\rmx e^{\hT^\rmx} + 
e^{-\hT^{\rmx\rmy}}\tH_{\rmx\rmy}e^{\hT^{\rmx\rmy}}
\end{align}
where summation over $\rmx$ and $\rmy$ is implied.
Here we have introduced the internally contracted, partially similarity transformed Hamiltonians
\begin{equation}\label{meffH}
\tH_\rmx =
\hH_{\rmx} + 
\left[
e^{-\hT^\rmy}e^{-\hT^{\rmx\rmy}}
(\hH_{\rmy}+\hH_{\rmx\rmy})
e^{\hT^{\rmx\rmy}}e^{\hT^\rmy}
\right]_{\rmx} 
\end{equation}
and
\begin{equation}\label{deffH}
\tH_{\rmx\rmy} = 
\hH_{\rmx\rmy} + 
\left[
e^{-\hT^\rmx}
\hH_{\rmx}
e^{\hT^\rmx}
+
e^{-\hT^\rmy}
\hH_{\rmy}
e^{\hT^\rmy}
\right]_{\rmx\rmy}
\end{equation}
by expanding Eq. \ref{fef2hbar1} then collecting terms either in the monomer
($\rmx$) or dimer ($\rmx\rmy$) spaces.  Here the brackets denote
$\left[\dots\right]_\rmx\in\cH_\rmx$ and
$\left[\dots\right]_{\rmx\rmy}\in\cH_\rmx\otimes\cH_\rmy$ with all terms within
the brackets not a member of the appropriate final space are to be internally
contracted away.  The bracket quantities are further defined to contain at least
one non-trivial vertex, so that 
\begin{equation}
\bra{\phi^{\rmx}_0}\left[\dots\right]_\rmx\ket{\phi^{\rmx}_0}=
\bra{\phi^{\rmx\rmy}_0}\left[\dots\right]_{\rmx\rmy}\ket{\phi^{\rmx\rmy}_0} = 0.
\end{equation}
Equations \ref{meffH} and \ref{deffH} represent a very powerful restructuring of the Hamiltonian by embedding the effective field of the surrounding system into a compact set of fragment Hamiltonians, which will be shown to advantage later.

\subsection{Fragment Effective Field Coupled-Cluster Perturbation Theory}

The general coupled-cluster perturbation theory
equations\cite{bartlett2010,byrd2015-a} and similarity transformed
Hamiltonian\cite{byrd2015-c} have been derived and discussed previously so we
will forgo a lengthy discussion.  Throughout we employ the particle rank conserving
partitioning of the Hamiltonian,
\begin{equation}\label{partitionedH}
\hH = \hH_0 + \alpha \hV\left.\right|_{\alpha\rightarrow 1},
\end{equation}
with
\begin{align}
\label{partitioning0}
\hH_0 &= \hF^{[0]} + \hW^{[0]} \\
\label{partitioningV}
\hV   &= \hF^{[\pm 1]} + \hW^{[\pm 1]} + \hW^{[\pm 2]}
\end{align}
where $\hF$ and $\hW$ are the one- and two-electron Hamiltonian operators and the $[n]$ superscript denotes that the operator changes particle rank from right to left by $n$.\footnote{As an example, the occupied-virtual Fock matrix element $f^a_i$ changes particle rank by 1 and so is represented as $\hF^{[+1]}$.}
We have previously derived the general CCPT similarity transformed Hamiltonian to arbitrary order.\cite{byrd2015-c} For convenience we give the first three orders of $\bH$
\begin{align}
\label{ccpth0}
\bH^{(0)} &= \hH_0, \\
\label{ccpth1}
\bH^{(1)} &= \left[\hH_0,\hT^{(1)}\right] + \hV, \\
\label{ccpth2}
\bH^{(2)} &= 
\left[\hH_0,\hT^{(2)}\right] + 
\left[\left[\hH_0,\hT^{(1)}\right],\hT^{(1)}\right] +
\left[\hV,\hT^{(1)}\right],
\end{align}
the first-order CCPT amplitude equations\cite{bartlett2010,byrd2015-c}
\begin{align}
%\label{1pr0wfDoubles}
%\bra{\phi_1}\bH^{(1)}\ket{\phi_0} =
%\bra{\phi_1} (\hF^{[0]} + \hW^{[0]})\hT_1 + \hF^{[+1]} \ket{\phi_0}= 0,\\
\label{2pr0wfDoubles}
%\bra{\phi_2}\bH^{(1)}\ket{\phi_0} = 
\bra{\phi_2} (\hF^{[0]} + \hW^{[0]})\hT_2 + \hW^{[+2]} \ket{\phi_0}= 0,
\end{align}
and the second-order CCPT correlation energy\cite{bartlett2010,byrd2015-c}
\begin{equation}\label{ccptE}
\bra{\phi_0}\bH^{(2)}\ket{\phi_0} - \bra{\phi_0}\hH\ket{\phi_0} 
= \Delta E^{(2)}_{\rm CC}.
\end{equation}

The FEF-CCPT wavefunction and correlation energy can be readily obtained now by using the above CCPT equations (Eqs. \ref{2pr0wfDoubles} and \ref{ccptE}) with the fragment cluster amplitudes and effective Hamiltonians (Eqs. \ref{clusterT}, \ref{meffH} and \ref{deffH} respectively).  Performing the substitutions the FEF-CCPT amplitude equations are
\begin{align}
\label{monDoubles}
\bra{\phi^\rmx_2}\bH^{(1)}_\rmx \ket{\phi^\rmx_0} =
\bra{\phi^\rmx_2} 
\tH^{[0]}_\rmx \hT^{\rmx(1)}_2 + \tH^{[+2]}_\rmx \ket{\phi^\rmx_0} &= 0 \\
\label{dimDoubles}
\bra{\phi^{\rmx\rmy}_2}\bH^{(1)}_{\rmx\rmy}\ket{\phi^{\rmx\rmy}_0} = 
\bra{\phi^{\rmx\rmy}_2} 
\tH^{[0]}_{\rmx\rmy} \hT^{{\rmx\rmy}(1)}_2 + \tH^{[+2]}_{\rmx\rmy} \ket{\phi^{\rmx\rmy}_0} &= 0.
\end{align}
Similarly, the correlation energy is
\begin{align}
\label{EFEF2}
E^{(2)}_{\rm corr} =&
\bra{\phi^\rmx_0}\bH^{(2)}_\rmx \ket{\phi^\rmx_0}
+
\bra{\phi^{\rmx\rmy}_0}\bH^{(2)}_{\rmx\rmy}\ket{\phi^{\rmx\rmy}_0}\\
=&
E^{(2)}_{\rm M2} + E^{(2)}_{\rm D2}
\end{align}
with the convenient decomposition of the energy defined by
\begin{align}
\label{EM2}
E^{(2)}_{\rm M2} &=
\sum_\rmx 
\bra{\phi^\rmx_0} \hW^{[-2]}_\rmx\hT^{\rmx(1)}_2\ket{\phi^\rmx_0} \\
\label{ED2}
E^{(2)}_{\rm D2} &=
\sum_{\rmy\ne\rmx}
\bra{\phi^{\rmy\rmx}_0} \hW^{[-2]}_{\rmx\rmy} \hT^{\rmx\rmy(1)}_2\ket{\phi^{\rmy\rmx}_0}.
\end{align}
By partially performing the coupled-cluster similarity transformation within the monomer and dimer spaces, the contributions of the surrounding system to the wavefunction of any particular monomer are completely encapsulated within the $\tH$ effective Hamiltonian.   An effective Hamiltonian representation of molecular cluster perturbation theory allows for an embedding  interpretation of FEF-CCPT and ready application within canonical CCPT methodologies as is illustrated by Eqs. \ref{monDoubles}-\ref{EFEF2}.

\subsection{Correction to Configuration Interaction Singles using FEF-CCPT}
{\color{blue}

\begin{align}
\label{cismonDoubles}
\bra{\phi^\rmx_2}\bH^{(1)}_\rmx \left(1+\hat{C}_{\rmx}(k)\right)\ket{\phi^\rmx_0}_L 
%=\bra{\phi^\rmx_2} \tH^{[0]}_\rmx \hT^{\rmx(1)}_2 + \tH^{[+2]}_\rmx \left(1+\hat{C}_1(k)\right)\ket{\phi^\rmx_0}_L 
&= 0 \\
\label{cisdimDoubles}
\bra{\phi^{\rmx\rmy}_2}\bH^{(1)}_{\rmx\rmy}\left(1+\hat{C}_{\rmx}(k)+\hat{C}_{\rmy}(k)\right)\ket{\phi^{\rmx\rmy}_0}_L 
%= \bra{\phi^{\rmx\rmy}_2} 
%\tH^{[0]}_{\rmx\rmy} \hT^{{\rmx\rmy}(1)}_2 + \tH^{[+2]}_{\rmx\rmy} \left(1+\hat{C}_1(k)\right)\ket{\phi^{\rmx\rmy}_0}_L 
&= 0.
\end{align}

\begin{align}
\label{CISEM2}
E^{(2)}_{\rm M2} &=
\sum_\rmx 
\bra{\phi^\rmx_0} \left(1+\hat{C}_{\rmx}^\dag(k)\right) \bH^{(2)}_\rmx \left(1+\hat{C}_{\rmx}(k)\right)\ket{\phi^\rmx_0}_L \\
\label{CISED2}
E^{(2)}_{\rm D2} &=
\sum_{\rmy\ne\rmx}
\bra{\phi^{\rmy\rmx}_0} \left(1+\hat{C}^\dag_{\rmx}(k)+\hat{C}^\dag_{\rmy}(k)\right)\bH^{(2)}_{\rmx\rmy}\left(1+\hat{C}_{\rmx}(k)+\hat{C}_{\rmy}(k)\right)\ket{\phi^{\rmy\rmx}_0}_L.
\end{align}

}

\subsection{Equation-of-Motion with FEF-CCPT}

{\color{blue}
The standard EOM-CCSD equations are 
\begin{equation}\label{eomccpt}
%\bra{\phi_0}\hLe(k')\left[\bH^{\lbrace 2\rbrace},\hR(k)\right]\ket{\phi_0} = \omega_k \delta_{k'k},
\bra{\phi_0}\hLe(k')\left[\bH,\hR(k)\right]\ket{\phi_0} = \omega_k \delta_{k'k},
\end{equation}
with the excitation and de-excitation operators are defined up to double excitations as
\begin{align}
\label{eomR}
\hR(k) &= r_0(k) + \hR_1(k) + \hR_2(k)\\
& = r_0(k) +
\sum_{ai} r^{a}_{i}(k)\hat{a}^\dag\hat{i} +
\frac{1}{4} \sum_{abij}r^{ab}_{ij}(k)\hat{a}^\dag\hat{i} \hat{b}^\dag \hat{j}\\
\label{eomL}
\hLe(k) &= \hLe_1(k) + \hLe_2(k) \\
& =
\sum_{ai} \ell^{i}_{a}(k)\hat{i}^\dag\hat{a} +
\frac{1}{4} \sum_{abij} \ell^{ij}_{ab}(k)\hat{i}^\dag\hat{a}\hat{j}^\dag\hat{b},
\end{align}
where $\hLe$ and $\hR$ are biorthogonal
\begin{equation}
\bra{\phi_0}\hLe(k)\hR(k')\ket{\phi_0}\equiv \delta_{kk'}.
\end{equation}
We derived previously the corresponding EOM-CCPT(n) equations using the
similarity transformed Hamiltonian $\bH^{\lbrace n\rbrace}$ which is {\it
inclusive} (throughout $(n)$ denotes exclusive and $\lbrace n\rbrace$ denotes
inclusive) of all orders up to
$n$,
\begin{equation}\label{incHB}
\bH^{\lbrace n\rbrace} = \hH_0 + \bH^{(1)} + \cdots + \bH^{(n)}.
\end{equation}
Using Eq. \ref{incHB} the EOM-CCPT(2) defining equation is
\begin{equation}\label{eomccpt}
\bra{\phi_0}\hLe(k')\left[\bH^{\lbrace 2\rbrace},\hR(k)\right]\ket{\phi_0} =
\omega_k \delta_{k'k}.
\end{equation}
The MCPT(2) inclusive similarity transformed Hamiltonian is, using 
Eq. \ref{fef2hbar1} inserted into Eqs.
\ref{ccpth0}-\ref{ccpth2},
\begin{equation}
\label{xFEFSTH}
\bH^{\lbrace 2\rbrace}_\rmx =
\tH_\rmx +
\left[\tH^{[0]}_\rmx,\hT^{\rmx(1)}\right] +   
\left[\tH^{[\pm 1]}_\rmx,\hT^{\rmx(1)}\right] +
\left[\tH^{[\pm 2]}_\rmx,\hT^{\rmx(1)}\right]
\end{equation}
and
\begin{multline}
\label{xyFEFSTH}
\bH^{\lbrace 2\rbrace}_{\rmx\rmy} =
\tH_{\rmx\rmy} +
\left[\tH^{[0]}_{\rmx\rmy},\hT^{\rmx\rmy(1)}\right] + \\
\left[\tH^{[\pm 1]}_{\rmx\rmy},\hT^{\rmx\rmy(1)}\right] +
\left[\tH^{[\pm 2]}_{\rmx\rmy},\hT^{\rmx\rmy(1)}\right].
\end{multline}
The EOM-FEF-CCPT(2) equations is then
\begin{equation}\label{eomfef}
\bra{\phi_0}\hLe(k')\left[
(\bH^{\lbrace 2\rbrace}_{\rmx} + \bH^{\lbrace 2\rbrace}_{\rmx\rmy}),
\hR(k)\right]\ket{\phi_0} =
\omega_k \delta_{k'k}.
\end{equation}
Partitioning Eqs. \ref{eomR} and \ref{eomL} into fragment subspaces, analogous to Eq. \ref{clusterT}, gives
\begin{align}
\label{eomFR}
\hR(k) &= \sum_\rmx \hR^\rmx(k) + \sum_{\rmy\ne\rmx} \hR^{\rmx\rmy}(k)\\
\label{eomFL}
\hLe(k) &= \sum_\rmx \hLe^\rmx(k) + \sum_{\rmy\ne\rmx} \hLe^{\rmx\rmy}(k).
\end{align}
By expanding Eqs. \ref{eomFR} and \ref{eomFL} in Eq. \ref{eomccpt} the inter-
and intra-fragment EOM-FEF-CCPT(2) equations are
\begin{align}
\label{fef-mee}
\bra{\phi_0}\hLe^\rmx(k')\left[\bH^{\lbrace 2\rbrace}_\rmx,\hR^\rmx(k)\right]\ket{\phi_0} 
&+  \\
\label{fef-fret}
\bra{\phi_0}\hLe^\rmy(k')\left[\bH^{\lbrace 2\rbrace}_{\rmx\rmy},\hR^\rmx(k)\right]\ket{\phi_0} 
&+ \rmx \rightleftharpoons \rmy + \\
\label{fef-fission}
\bra{\phi_0}\hLe^\rmy(k')\left[\bH^{\lbrace 2\rbrace}_{\rmx\rmy},\hR^{\rmx\rmy}(k)\right]\ket{\phi_0} 
&+ \rmx\rmy \rightleftharpoons \rmy + \\
\label{fef-xyfret}
\bra{\phi_0}\hLe^{\rmx\rmy}(k')\left[\bH^{\lbrace 2\rbrace}_{\rmx\rmy},\hR^{\rmx\rmy}(k)\right]\ket{\phi_0} 
&= \omega_k.
\end{align}
The EOM-FEF-CCPT(2) equations as expanded above offer physical interpretation of
contributions to the computed spectra.
The intra-fragment excited states computed from Eq. \ref{fef-mee} correspond to
the embedding field perturbing the fragment $\rmx$, while Eqs.
\ref{fef-fret}-\ref{fef-xyfret} correspond to (exchangeless) excitation transport between
fragments (fluorescence resonance energy transfer).  In this work we are
focusing only on the intra-fragment excited state spectra and leave a detailed
examination of the excitation transport equations for a later manuscript.

The matrix elements of Eq. \ref{xFEFSTH} are identical to those worked out elsewhere,\cite{stanton1995,nooijen1995,gwaltney1996,byrd2015-c} with the following additional cross monomer terms
\begin{align}
\label{MPij}
\cH^{i}_{i_1} =& \dots + \sum_{\rmy\ne\rmx}\left(
    v^{ij_1}_{i_1b_1} t^{b_1}_{j_1} +
    v^{ij_1}_{a_1b_1} t^{a_1b_1}_{i_1j_1}
    \right)
,\\
\label{MPab}
\cH^{a}_{a_1} =& \dots + \sum_{\rmy\ne\rmx}\left(
    v^{aj_1}_{a_1b_1} t^{b_1}_{j_1} - 
    v^{i_1j_1}_{a_1b_1} t^{ab_1}_{i_1j_1}
    \right)
,\\
\label{MPia}
\cH^{i}_{a} =& \dots + \sum_{\rmy\ne\rmx}
    \bv^{ij_1}_{ab_1} t^{b_1}_{j_1}
,\\
\label{mPiabj}
\cH^{ia_1}_{ai_1} =& \dots + P(ii_1)P(aa_1)\sum_{\rmy\ne\rmx}
    v^{ij_1}_{ab_1} t^{a_1b_1}_{i_1j_1} 
,\\
\label{MPiajk} 
\cH^{ia}_{i_1i_2} =&  \dots + P(ii_1)\sum_{\rmy\ne\rmx}
    v^{ij_1}_{i_1b_1} t^{a_1b_1}_{i_2j_1} 
,\\
%\nonumber
\label{MPabci}
\bH^{aa_1}_{a_2i} =& 
    \dots - P(aa_1)\sum_{\rmy\ne\rmx}
    v^{aj_1}_{a_2b_1} t^{a_1b_1}_{ij_1}
%\right)
,
\end{align}
}

\end{document}